\documentstyle[12pt]{article}
\newcommand{\be}{\begin{equation}}
\newcommand{\ee}{\end{equation}}
\newcommand{\ba}{\begin{eqnarray}}
\newcommand{\ea}{\end{eqnarray}}

\newcommand{\bi}{\bibitem}
\begin{document}
\begin{center}
{\bf{ ABOUT DUALITY AND KILLING TENSORS}}
\end{center}
\begin{center}
 Dumitru Baleanu\footnote{Permanent Address: Institute for Space Sciences,P.O.BOX, MG-23, R 76900
Magurele-Bucharest, Romania}
\footnote{e-mail:baleanu@thsun1.jinr.ru,baleanu@venus.nipne.ro}
\end{center}
\begin{center}
JINR, Bogoliubov Laboratory of Theoretical Physics
\end{center}
\begin{center}
{141980 Dubna,Moscow Region, Russia}
\end{center}

\begin{abstract}
  In this paper the
isometries of the dual space
were investigated.The dual structural equations of a Killing tensor of order two
 were  found .
The flat space case was analyzed in details.

\end{abstract}

\newpage\setcounter{page}2
\section{Introduction}
Killing tensors are indispensable tools in the quest for exact
solutions in many branches of general relativity as well as classical
mechanics \cite{gib}.
 Killing tensors are important for solving the
equations of motion in particular space-times.  The notable example here
is the Kerr metric which admits a second rank Killing tensor
\cite{gib}. \\
The Killing tensors give rise to new exact solutions in perfect fluid Bianchi
and Katowski-Sachs cosmologies as well in inflationary models with a scalar
field sources \cite{ros1}.
 Recently the Killing tensors of third rank in $(1+1)$ dimensional
geometry were investigated and classified \cite{ros}.
In a geometrical setting , symmetries are
connected with isometries associated with Killing vectors, and more
generally, Killing tensors on the configuration space of the system.An
example is the motion of a point particle in a space with isometries
\cite{hol}, which is a physicist's way of studying the geodesic
structure of a manifold.  Any  symmetrical tensor $K_{\alpha\beta}$
satisfying the condition
\be\label{tk} K_{(\alpha\beta;\gamma)}=0 ,\ee
 is called a Killing tensor. Here the parenthesis denotes a full
 symmetrization with all indices and  coma denotes a covariant
derivative.$K_{\alpha\beta}$ will be called redundant if it is equal to
some linear combination with constants coefficients of the metric
tensor $g_{\alpha\beta}$ and of the form $S_{(\alpha}B_{\beta)}$ where
$A_{\alpha}$ and $B_{\beta}$ are Killing vectors. For
any Killing vector $K_{\alpha}$ we have \cite{hau}
\be\label{veck}
K_{\beta;\alpha}=\omega_{\alpha\beta}=-\omega_{\beta\alpha},
\ee
\be\label{veck1}
\omega_{\alpha\beta;\gamma}=R_{\alpha\beta\gamma\delta}K^{\delta}.
\ee
The equations (\ref{veck}) and (\ref{veck1}) may be regarded as a
system of linear homogeneous first-order equations in the components
$K_{\alpha\beta}$ and $\omega_{\alpha\beta}$. In \cite{hau} equations
analogous to the above ones for a Killing vector were derived for
$K_{\alpha\beta}$.
Recently Holten \cite{holi} has presented a theorem concerning the
reciprocal relation between two local geometries described by metrics
which are Killing tensors with respect to one another.
In this paper the geometric duality was presented
and the structural equations of $K_{\alpha\beta}$ were analyzed.\\
The plan of this paper is as follows.
In Sec.2 the geometric duality is presented.
In Sec.3 the structural equations of $K_{\alpha\beta}$ are investigated.
Our comments and concluding remarks are presented in Sec. 4.

\section{ Geometric duality}

 Let us consider  that the space with a metric $g_{\alpha\beta}$
admits a Killing tensor field $K_{\alpha\beta}$.

As it is well known the equation of motion
of a particle on a geodesic is derived from the action  \cite{bal}
\be
S=\int{d\tau({1\over 2} g_{\alpha\beta}\dot{x^{\alpha}}{\dot x^{\beta}})}.
\ee
The Hamiltonian is constructed in the following form
 $H={1\over 2}g_{\alpha\beta}p^{\alpha}p^{\beta}$ and
the Poisson brackets are
\be
\{x_{\alpha}, p^{\beta}\}=\delta_{\alpha}^{\beta}.
\ee

The equation of motion for a phase space function $F(x,p)$ can be computed
from the Poisson brackets
with the Hamiltonian
\be
{\dot F}=\{F,H\},
\ee
where ${\dot F}={dF\over d\tau}$.
From the covariant component $K_{\alpha\beta}$ of the Killing tensor we
can construct a constant of motion K
\be
K={1\over 2}K_{\alpha\beta}p^{\alpha}p^{\beta}.
\ee
We can easy verify that
\be
\{H,K\}=0.
\ee

The formal similarity between the constants of motion $H$ and $K$ ,
and the symmetrical  nature of the condition implying the existence
of the Killing tensor amount to a reciprocal relation between two
different models: the model with Hamiltonian $H$ and constant of motion $K$,
and a model with constant of motion $H$ and Hamiltonian $K$.The relation
between the two models has a geometrical interpretation:it implies that
if $K_{\alpha\beta}$ are the contravariant components of a Killing tensor with
respect to the metric $g_{\alpha\beta}$, then $g_{\alpha\beta}$
must represent a
Killing tÿÿsor with respect to the metric defined by $K_{\alpha\beta}$.
 When $K_{\alpha\beta}$ has an inverse we interpret it as the
metric of another space and we can define the associated
Riemann-Christoffel connection $\hat \Gamma_{\alpha\beta}^{\lambda}$ as
usual through the metric postulate  ${\hat
D}_{\lambda}K_{\alpha\beta}=0 $.  Here ${\hat D}$ represents the
covariant derivative with respect to $K_{\alpha\beta}$.

The   relation between connections ${\hat \Gamma^{\mu}_{\alpha\beta}}$
and $\Gamma^{\mu}_{\alpha\beta}$ is \cite{bal1}

\be\label{cone}
{\hat \Gamma^{\mu}_{\alpha\beta}}=\Gamma^{\mu}_{\alpha\beta}-
K^{\mu\delta}D_{\delta}K_{\alpha\beta}.
\ee

As it is well known
for a given metric
$g_{\alpha\beta}$ the conformal transformation is defined as
 ${\hat
g_{\alpha\beta}}=e^{2U(x)}g_{\alpha\beta}$ and
  the  relation between the corresponding connections is
\be\label{game}
\hat\Gamma^{\lambda}_{\alpha\beta}=\Gamma^{\lambda}_{\alpha\beta}
 +2\delta^{\lambda}_{(\alpha}U_{\beta)^{'}}- g_{\alpha\beta}U^{'\lambda},
\ee
where $U^{'\lambda}={dU^{\lambda}\over dx}$.
 After some  calculations we conclude that the dual transformation  (\ref{cone})
is not a conformal transformation .\\
  For this reason  is interesting  to investigate
when the   manifold  and its dual  have the same isometries.
 Let us  denote by
$\chi_{\alpha}$  a Killing vector corresponding to $g_{\alpha\beta}$ and by
$\hat\chi_{\alpha}$ a Killing vector corresponding to
$K_{\alpha\beta}$.\\

{\bf Proposition }\\

The manifold  and its  dual
 have the same
Killing vectors iff
\be\label{vec}
 (D_{\delta}K_{\alpha\beta})\hat\chi^{\delta}=0.
\ee

{\bf Proof.}\\
Let us consider $\chi_{\sigma}$
a  vector satisfies
\be\label{killing}
D_{\alpha}\chi_{\beta} +D_{\beta}\chi_{\alpha}=0.
\ee

Using (\ref{cone})  the  corresponding  dual Killing vector's
equations are

\be\label{kildual}
D_{\alpha}\hat\chi_{\beta} +
D_{\beta}\hat\chi_{\alpha}
+2K^{\delta\sigma}(D_{\delta}K_{\alpha\beta}){\hat\chi_{\sigma}}=0.
\ee
 Let us  suppose that  $\hat\chi_{\alpha}=\chi_{\alpha}$, then using (\ref{killing})
and (\ref{kildual}) we obtain
\be
(D_{\delta}K_{\alpha\beta})\hat\chi^{\delta}=0.
\ee
 Conversely if we suppose that (\ref{vec}) holds , then from (\ref{kildual})
 we  can deduce  immediately that  $\chi_{\alpha}=\hat\chi_{\alpha}$.
{\bf q.e.d.}

\section {The structural
equations}

The following two vectors play roles analogous to that of a bivector
$\omega_{\alpha\beta}$
\be\label{prima}
L_{\alpha\beta\gamma}=K_{\beta\gamma;\alpha}- K_{\alpha\gamma;\beta},
\ee
\be\label{doua}
M_{\alpha\beta\gamma\delta}={1\over 2}(L_{\alpha\beta[\gamma;\delta]} +L_{\gamma\delta[\alpha;\beta]}).
\ee
The properties of the tensors $L_{\alpha\beta\gamma}$ and
 $M_{\alpha\beta\gamma\delta}$ were derived in
\cite{hau}.$M_{\alpha\beta\gamma\delta}$ has the same symmetries as
the Riemannian tensor and the covariant derivatives of
$K_{\alpha\beta}$ and $L_{\alpha\beta\gamma}$ satisfy relations
reminiscent of those satisfied  by Killing vectors.\\
From (\ref{prima})
and (\ref{doua}) we found  that
$M_{\alpha\beta\gamma\delta}={1\over 2}(K_{\beta\gamma;(\alpha\delta)} +K_{\alpha\delta;(\beta\gamma)}-K_{\alpha\gamma;(\beta\delta)}
-K_{\beta\delta;(\alpha\gamma)})$  and
$M_{\alpha\beta\gamma\delta} +M_{\gamma\alpha\beta\delta} +M_{\beta\gamma\alpha\delta}=0$ \cite{hau}.\\
  We will investigate now  the  dual structural
equations.Let us  define a tensor $H^{\mu}_{\alpha\beta}$ as
\be\label{hasuri}
H^{\mu}_{\alpha\beta}= {\hat\Gamma^{\mu}_{\alpha\beta}}-
 \Gamma^{\mu}_{\alpha\beta}.
\ee
 Taking into account (\ref{cone}) we  found

\be\label{nhas}
 H^{\delta}_{\alpha\beta}K_{\delta\gamma}= -
D_{\gamma}K_{\alpha\beta}.
\ee

Using  (\ref{prima}) and (\ref{nhas})
we obtain
\be\label{cin}
L_{\alpha\beta\gamma}=
 H^{\delta}_{\alpha\gamma}K_{\delta\beta}
- H^{\delta}_{\beta\gamma}K_{\delta\alpha}.
\ee
From (\ref{cin}) we conclude that
$L_{\alpha\beta\gamma}$ looks like an angular momentum.
This
result is in agreement with those presented in \cite{hau}.
Taking into account  (\ref{doua}) and (\ref{nhas})
the expression of $M_{\alpha\beta\gamma\delta}$  becomes\\

\ba\label{m}
&M_{\alpha\beta\gamma\delta}&={K_{\sigma\alpha}\over 2}[-D_{\delta}H^{\sigma}_{\beta\gamma}+ D_{\gamma} H^{\sigma}_{\beta\delta} +H^{\sigma}_{\theta\gamma}H^{\theta}_{\delta\beta}
- H^{\theta}_{\gamma\beta}H^{\sigma}_{\theta\delta}]\cr
&+&
{K_{\sigma\beta}\over 2}[D_{\delta}H^{\sigma}_{\alpha\gamma}- D_{\gamma}H^{\sigma}_{\alpha\delta}+H^{\theta}_{\delta\alpha}H^{\sigma}_{\theta\gamma}
- H^{\theta}_{\gamma\alpha}H^{\sigma}_{\theta\delta}]\cr
&+& {K_{\sigma\gamma}\over 2}[-D_{\beta} H^{\sigma}_{\delta\alpha} + D_{\alpha}H^{\sigma}_{\delta\beta} -H^{\theta}_{\beta\delta}H^{\sigma}_{\theta\alpha}
+ H^{\theta}_{\alpha\delta}H^{\sigma}_{\theta\beta}]\cr
&+& {K_{\sigma\delta}\over 2}[-D_{\alpha}H^{\sigma}_{\gamma\beta} +D_{\beta}H^{\sigma}_{\gamma\alpha}
+ H^{\theta}_{\beta\gamma}H^{\sigma}_{\theta\alpha} -H^{\theta}_{\alpha\gamma}H^{\sigma}_{\theta\beta}].
\ea

The general solution of eq. (\ref{tk}) in the flat space case
has the form
\be\label{k}
K_{\beta\gamma}=s_{\beta\gamma}+{2\over 3}B_{\alpha(\beta\gamma)}x^{\alpha} +{1\over 3}A_{\alpha\beta\gamma\delta}x^{\alpha}x^{\delta}.
\ee
Here $s_{\beta\gamma}$, $B_{\alpha\beta\gamma}$ and $A_{\alpha\beta\gamma\delta}$ are constant
tensors having the same symmetries as $K_{\beta\gamma}$,$L_{\alpha\beta\gamma}$ and $M_{\alpha\beta\gamma\delta}$
respectively.
Using (\ref{m}) and (\ref{k}) the expression of
$M_{\alpha\beta\gamma\delta}$ becomes
\be\label{mi}
M_{\alpha\beta\delta\gamma}={1\over
2}(K_{\sigma\alpha}R^{'\sigma}_{\beta\gamma\delta}
+K_{\sigma\beta}R^{'\sigma}_{\alpha\delta\gamma} +
K_{\sigma\gamma}R^{'\sigma}_{\delta\alpha\beta}
+K_{\sigma\delta}R^{'\sigma}_{\gamma\beta\alpha}), \ee
where
\be\label{r}
R^{'\beta}_{\nu\rho\sigma} =
H^{\beta}_{\nu\sigma,\rho} - H^{\beta}_{\nu\rho,\sigma}
+ H^{\alpha}_{\nu\sigma}H^{\beta}_{\alpha\rho}- H^{\alpha}_{\nu\rho} H^{\beta}_{\alpha\sigma}.
\ee

{} From (\ref{r}) we conclude that  $R^{'\beta}_{\nu\rho\sigma}$ looks like as
 the curvature tensor $R^{\beta}_{\nu\rho\sigma}$ \cite{dir} .\\
The next
step is to investigate the form of $M_{\alpha\beta\gamma\delta}$ on
a curved space.
After tedious calculations we found
the  expression of $M_{\alpha\beta\gamma\delta}$ as
\ba
&M_{\alpha\beta\gamma\delta}&=K_{\sigma\alpha}R^{'\sigma}_{\beta\gamma\delta} +K_{\sigma\beta}R^{'\sigma}_{\alpha\delta\gamma} +K_{\sigma\gamma}R^{'\sigma}_{\delta\alpha\beta} +K_{\sigma\delta}R^{'\sigma}_{\gamma\beta\alpha}\cr
&-& K_{\sigma\chi}(-H^{\chi}_{\beta\delta}G^{\sigma}_{\alpha\gamma} +H^{\chi}_{\alpha\delta} G^{\sigma}_{\beta\gamma} +
H^{\chi}_{\beta\gamma}G^{\sigma}_{\alpha\delta} - H^{\chi}_{\alpha\gamma}G^{\sigma}_{\beta\delta})\cr
&+& H^{\sigma}_{\beta\delta}K_{\alpha\gamma,\sigma}+H^{\sigma}_{\alpha\gamma}K_{\beta\delta,\sigma}
-H^{\sigma}_{\beta\gamma}K_{\alpha\delta,\sigma}-
H^{\sigma}_{\alpha\delta}K_{\beta\gamma,\sigma},
\ea
where
$G^{\mu}_{\alpha\beta}=- H^{\mu}_{\alpha\beta}
+\Gamma^{\mu}_{\alpha\beta}$.
 $M_{\alpha\beta\gamma\delta}$
 has the form (\ref{mi})
for any curve $\gamma(\tau)$ belonging to manifold \cite{eisen}.

Using  (\ref{hasuri}) and (\ref{nhas}) we
found the dual expressions of $L_{\alpha\beta\gamma\delta}$ and $M_{\alpha\beta\gamma\delta}$ as
\ba
&{\hat L_{\alpha\beta\gamma}}&= {\hat
D_{\alpha}}g_{\beta\gamma}- {\hat
D_{\beta}}g_{\alpha\gamma}=-H^{\delta}_{\alpha\gamma}g_{\delta\beta}+
H^{\delta}_{\beta\gamma}g_{\delta\alpha}\cr
&{\hat M_{\alpha\beta\gamma\delta}}&={1\over 2}({\hat L_{\alpha\beta[\gamma;\delta]}}
 +{\hat L_{\gamma\delta[\alpha;\beta]}})
=-g_{\sigma\alpha}R^{'\sigma}_{\beta\gamma\delta} -g_{\sigma\beta}R^{'\sigma}_{\alpha\delta\gamma} -g_{\sigma\gamma}R^{'\sigma}_{\delta\alpha\beta} -g_{\sigma\delta}R^{'\sigma}_{\gamma\beta\alpha}\cr
&-& g_{\sigma\chi}(H^{\chi}_{\beta\delta}{\hat G^{\sigma}_{\alpha\gamma}} -H^{\chi}_{\alpha\delta}{\hat G^{\sigma}_{\beta\gamma}} +
- H^{\chi}_{\beta\gamma}{\hat G^{\sigma}_{\alpha\delta}} + H^{\chi}_{\alpha\gamma}{\hat G^{\sigma}_{\beta\delta}})\cr
&-& H^{\sigma}_{\beta\delta}g_{\alpha\gamma,\sigma}-H^{\sigma}_{\alpha\gamma}g_{\beta\delta,\sigma}
+H^{\sigma}_{\beta\gamma}g_{\alpha\delta,\sigma}
+H^{\sigma}_{\alpha\delta}g_{\beta\gamma,\sigma},
\ea
Here $\hat G^{\sigma}_{\alpha\delta}= H^{\sigma}_{\alpha\delta} +{\hat\Gamma^{\sigma}_{\alpha\delta}}$
and the semicolon denotes the dual covariant derivative .
 Taking into account (\ref{hasuri})
we found a new identity for $K_{\alpha\beta}$
\be\label{nd}
K^{\sigma}_{\beta} D_{\sigma}K_{\nu\lambda}
+K^{\sigma}_{\lambda} D_{\sigma}K_{\beta\nu}
+K^{\sigma}_{\nu} D_{\sigma}K_{\beta\lambda}=0. \ee
By duality we get from (\ref{nd}) the following identity  for $g_{\alpha\beta}$
\be\label{gdu} g^{\sigma}_{\beta}{\hat D}_{\sigma}g_{\nu\lambda}
+g^{\sigma}_{\lambda}{\hat D}_{\sigma}g_{\beta\nu}
+g^{\sigma}_{\nu}{\hat D}_{\sigma}g_{\beta\lambda}=0.
\ee
\section{Conclusions}

The geometric duality between
local geometry described by
$g_{\alpha\beta}$ and  the local geometry described by  Killing tensor
$K_{\alpha\beta}$ was presented.  We found  the  relation
between connections corresponding to $g_{\alpha\beta}$ and
$K_{\alpha\beta}$ respectively  and  we have shown that the dual
  transformation is not a conformal transformation.
The manifold  and its dual have the same isometries if
 $D_{\lambda}K_{\alpha\beta}=0$.
 We have shown  that $L_{\alpha\beta\gamma}$ looks like an angular momentum.
The  dual structural equations
 were analyzed and
the expressions of ${\hat L_{\alpha\beta\gamma}}$ and ${\hat M_{\alpha\beta\gamma\delta}}$
were calculated.For the flat space case the  general forms of $(L_{\alpha\beta\gamma},
{\hat L_{\alpha\beta\gamma}})$ and  $(M_{\alpha\beta\gamma\delta},  {\hat M_{\alpha\beta\gamma\delta}})$  were found.

\section{Acknowledgments}
I would like to thank TUBITAK for financial
support  and METU for the hospitality during the
working stage at Department of Physics.

\end{document}